\documentclass[aps, prb, twocolumn,floats, showpacs]{revtex4-1}
\usepackage{amsmath}
\usepackage{epsfig}
\usepackage{amsfonts}
\usepackage{amsbsy}
\usepackage{amssymb}
\usepackage{graphicx}
\usepackage{textcomp}
\usepackage{subfig}
\usepackage{xcolor}
\usepackage{mathrsfs}
\usepackage[colorlinks,linkcolor=red,citecolor=blue,filecolor=magenta,
urlcolor=cyan]{hyperref}

\begin{document}
\title{Quantum geometry of correlated many-body states}

\author{S. R. Hassan}
\email{shassan@imsc.res.in}
\affiliation{The Institute of Mathematical Sciences, C.I.T. Campus, Chennai 600 
113, India}
\affiliation{Homi Bhabha National Institute, Training School Complex, Anushakti Nagar, Mumbai 400094, India}
\author{R. Shankar}
\email{shankar@imsc.res.in}
\affiliation{The Institute of Mathematical Sciences, C.I.T. Campus, Chennai 600 
113, India}
\affiliation{Homi Bhabha National Institute, Training School Complex, Anushakti Nagar, Mumbai 400094, India}
\author{Ankita Chakrabarti}
\email{ankitac@imsc.res.in}
\affiliation{The Institute of Mathematical Sciences, C.I.T. Campus, Chennai 600 
113, India}
\affiliation{Homi Bhabha National Institute, Training School Complex, Anushakti Nagar, Mumbai 400094, India}

\date{\today}

\begin{abstract}
We provide a definition of the quantum distances of correlated many fermion
wave functions in terms of the expectation values of certain operators that we
call exchange operators. We prove that the distances satisfy the triangle
inequalities. We apply our formalism to the one-dimensional $t-V$ model, which
we solve numerically by exact diagonalisation. We compute the distance matrix
and illustrate that it shows clear signatures of the metal-insulator
transition.
\end{abstract}
\maketitle

\section{Introduction}

There is a growing realization that quantum geometry is a useful way of
characterizing many body ground states of interacting fermions in a periodic
potential \cite{haldanetalk2012,resta2011}. Quantum geometry is a very
general concept, valid for any quantum system. It is a characterization of the
states and hence is a way to describe the kinematics of quantum systems
\cite{mukunda1993}.  The inner product of the Hilbert space, which is the basis of
the physical interpretation of states, naturally defines a distance between two
states and a geometric phase associated with three states \cite{mukunda1993}.
The distances satisfy the triangle inequalities. The geometric phases satisfy
an additive law (detailed later). If there is a subspace of the Hilbert space,
parameterised by a set of variables, such that the distances and geometric
phases are smooth functions of the parameters, then they define a quantum
metric and the so called Berry curvature (BC) in the parameter space \cite{sriluckshmy2014,faye2014}.

In condensed matter systems, the relevence of quantum geometry was first 
pointed out in the pioneering paper by Thouless et. al. \cite{thouless1982},
where it was shown that, for non interacting electrons in a magnetic field
and a periodic potential, if the quasi-momenta were chosen as the parameters
parameterising the single particle states, then the Hall conductivity could
be identified with the Chern invariant which is the integral of the BC over 
the Brillouin zone (BZ).  It was later pointed out \cite{sundaram1999, haldanetalk2006,
Niurmp,haldane2004} that the BC can be identified with the so called anomalous velocity
discovered by Karplus and Luttinger \cite{Karplus}. Quantum distances and metric of
the single particle states were discussed by Marzari and Vanderbilt 
\cite{marzari1997} in the context of the spread functional of Wannier orbitals.

In a classic paper \cite{Walterkohn}, Walter Kohn had proposed that it is
possible to characterize insulators in terms of the structure of the ground
state alone. This idea was developed further by several others 
\cite {Resta1,RestaSorella,SouzaMartin, Resta2, Resta3,restalecture2015},
using quantum geometry to describe the structure of the ground state. In
particular, the localisation tensor was identified with the integral of the
quantum metric over the BZ \cite{ RestaSorella}. This tensor is
finite in the insulating phase and divergent in the metallic phase. Thus, this
line of work has led to a geometric theory of the insulating state. \cite{restalecture2015}

In all the work discussed above, the quantum distances between two
quasi-momenta, the geometric phase associated with three quasi-momenta and the
corresponding quantum metric and BC on the BZ is defined in terms of single
particle states and used to characterize the quantum geometry of mean-field
states. Can these concepts be meaningfully generalised to describe the geometry
of correlated states?  Global quantities, the integral of the BC over the BZ
(the Chern invariant) and the integral of the quantum metric over the BZ (the
localization tensor) can be defined in terms of the response of the system to
changes in the boundary conditions \cite{Niurmp}. However, this does not lead to
a defintion of the local quantities, namely the quantum distance between two
quasi-momenta and the geometric phase associated with three quasi-momenta. One
approach has been to define these quantities in terms of the zero frequency
limit of the Euclidean Green's function \cite{Shou-ChengZhang1, 
Shou-ChengZhang2, Dai1, Gurarie, PatricLee, Shou-ChengZhang3}.
However, this quantity is not a purely ground state property, but involves all
the single particle excitations.

The ``structure of the ground state", or any many-body state is completely
characterized by the static correlation functions. In this paper, we propose a
definition of the quantum distances of any many-body state in terms of static
correlation functions. The building blocks of many body states are single
particle states. The complete set of single particle states can be labelled by
some set of parameters that we refer to as the spectral parameters. These could
be the quasi-momenta in periodic systems or any other set of quantum numbers.
We show that quantum distances on the space of spectral parameters, can be
defined in terms of the expectation value of certain operators that we call the
exchange operators. Our definition reduces to the standard one in terms of
single particle states for mean field states. 

Our formalism yields non-trivial results even for partially filled single band
systems. Thus, unlike the single particle formalism, it is capable of probing
the quantum geometry of metallic phases as well as insulating ones. To
illustrate this, we apply our definition to a simple but non-trivial model, the
one-dimensional $t-V$ model. We solve the model (up to 18 sites) using exact
diagonalisation, compute the distance matrix and illustrate that it shows
clear signals of the metal-insulator transition.

The rest of this paper is organised as follows. In Section \ref{dgprev}, we
briefly review the basic concepts of quantum geometry. Section \ref{bimps}
describes our definition of the quantum distances characterizing many-particle
states in terms of the expectation value of products of exchange operators.
The triangle inequalities for our definition of quantum distances are proved in
Section \ref{tineq}.  Section \ref{efops} derives explicit expressions for the
exchange operators in terms of the fermion creation and annihilation operators.
The formalism is applied to study the quantum distances of the one-dimensional
$t-V$ model in Sections \ref{tv1}, \ref{edm} and \ref{ndm}. We discuss our
results and conclude in Section \ref{dc}.

\section{A brief review of quantum distances and geometric phases}
\label{dgprev}

In this section we briefly review the basic concepts of quantum geometry
and its application in condensed matter systems. 

In quantum theory, physical states are represented by rays in a Hilbert space.
All observable physical quantities are therefore functions on the space of 
rays, the projective Hilbert space. These can expressed in terms of the so
called Bargmann invariants \cite{mukunda1993,Bargmann}, which are constructed 
using the inner product as follows. A state $\psi$ can be represented by the 
pure state density matrix,
$\rho(\psi)\equiv\vert\psi\rangle\langle\psi\vert$, with 
${\rm tr}\rho(\psi)=1$. The second order Bargmann invariant associated with
two states, $(\psi_1,\psi_2)$ is defined as,
\begin{equation}
\label{b2def}
{\cal B}^2(\psi_1,\psi_2)\equiv{\rm tr}\left(\rho(\psi_1)\rho(\psi_2)\right).
\end{equation}
The $n^{\rm th}$ order Bargmann invariant associate with any ordered sequence 
of $n$ states,
$\left\{\psi\right\}\equiv\left(\psi_1,\psi_2,\dots,\psi_n\right)$, is defined
as,
\begin{equation}
\label{bndef}
{\cal B}^n\equiv{\rm tr}\left(\rho(\psi_1)\rho(\psi_2)\dots\rho(\psi_n)\right).
\end{equation}
The Bargmann invariants have a geometric interpretation in terms of quantum
distances and geometric phases \cite{Simon1993}. 

${\cal B}^2(\psi_1,\psi_2)$ defines a distance, $d(\psi_1,\psi_2)$, between
the two states, namely a segment in the projective Hilbert space.
\begin{equation}
\label{dpsi1psi2def}
d(\psi_1,\psi_2)\equiv\sqrt{1-\left({\cal B}^2(\psi_1,\psi_2)\right)^\alpha}.
\end{equation}
This definition is consistent if, $d(\psi_i,\psi_j)$ satisfy the following 
properties,
\begin{eqnarray}
\label{dprop1}
d(\psi_i,\psi_i)&=&0\\
\label{dprop2}
d(\psi_i,\psi_j)&=&d(\psi_j,\psi_i)\\
\label{dprop3}
d(\psi_i,\psi_j)+d(\psi_j,\psi_k)&\ge&d(\psi_i,\psi_k).
\end{eqnarray}
The first two properties are obvious from the definition Eq.~\eqref{dpsi1psi2def}. 
For $\alpha\ge 0.5$, the definition in Eq.~\eqref{dpsi1psi2def} 
satisfies the triangle inequalities defined in Eq.~\eqref{dprop3}. 

The phase of the $n^{th}$ order invariant defines the the geometric phase
associated with the loop in the projective Hilbert space defined by the ordered
sequence of states, $\{\psi\}$. The ``loop" being defined as the union of the
segments, $(\psi_i,\psi_{i+1})$ with $\psi_{n+1}\equiv\psi_1$. This
identification is possible because the phases of the Bargmann invariants (by
construction) satisfy an additive law: if a loop can be expressed as a union of
several smaller loops, then the sum of the phases of the Bargmann invariants
associated with the smaller loops must equal to the phase of the full loop.
eg. consider four points in the projective Hilbert space
$(\psi_1,\psi_2,\psi_3,\psi_4)$.  Denote the phases of the loops consisting of
$n$ points ($n\le 4$), that can be constructed from these four points as
$\Omega^{(n)}(\{\psi\}),~n=3,4$. Then, the additive law implies,
\begin{eqnarray}
\label{addlaw1}
\Omega(\psi_1,\psi_2,\psi_3,\psi_4)&=&
\Omega(\psi_1,\psi_2,\psi_3)+\Omega(\psi_1,\psi_3,\psi_4)\\
&=&
\label{addlaw2}
\Omega(\psi_1,\psi_2,\psi_4)+\Omega(\psi_2,\psi_3,\psi_4).
\end{eqnarray}
As mentioned earlier, the properties in Eqs.~\eqref{addlaw1},\eqref{addlaw2}
follow trivially from the definition in Eq.~\eqref{bndef}.

The quantum geometry reviewed above has been applied to examine the structure
of the many-fermion states in periodic systems \cite{Niurmp}.
Consider an $N_B$ band tight-binding model. We denote the
single-particle Hamiltonian in the quasi-momentum space by an $N_B\times N_B$
matrix, $h(\mathbf k)$, where $\mathbf k$ takes values in the Brillouin zone.
Its spectrum is denoted by, $h(\mathbf k)u^n(\mathbf k)=\epsilon_n(\mathbf
k)u^n(\mathbf k)$. The single-particle states in the $n^{\rm th}$ band are
denoted by $\rho^n(\mathbf k)=u^n(\mathbf k)\left(u^n(\mathbf
k)\right)^\dagger$.

For every band of states, we can associate a Bargmann invariant with
every ordered sequence of $n$ points in the Brillouin zone, $\{\mathbf k\}
=(\mathbf k_1,\mathbf k_2,\dots,\mathbf k_n)$ using the definitions in Section
\ref{dgprev}. This yields a definition of a distance between any two points
in the Brillouin zone and a geometric phase associated with every loop in it. 

Basically, the single-particle states, $u^n(\mathbf k)$ define a mapping
from the Brillouin zone to the projective Hilbert space ($CP_{N_B-1}$ in
this case), $\mathbf k\rightarrow\rho^n(\mathbf k)$. The quantum 
distances and geometric phases in the Brillouin zone come from this map.
If the image of the Brillouin zone under this map is a smooth surface in the
projective Hilbert space, then it is possible to define a quantum metric and
curvature from the distances between infinitesimally seperated points and
the geometric phases of infinitesimal loops.

\section{Quantum distances for many-particle states}
\label{bimps}

In this section, we first consider mean field states and show that the
expressions for the Bargmann invariants defined in Eq.~\eqref{bndef} can
be written as the expectation values of the products of certain operators that
we call exchange operators. We then define quantum distances for correlated
states and prove that the definition satisfies the triangle inequality.
Finally, we discuss some difficulties in defining the geometric phases for the
correlated states.

\subsection{Mean field states}

We consider a system  of $N_B$ species of fermions on a $d$-dimensional,
lattice with $L^d$ unit cells. We denote the basis vectors by $\mathbf
e_\mu,~\mu=1,\dots,d$ and the sites by $\mathbf I=\sum_{\mu=1}^di_\mu\mathbf
e_\mu,~i_\mu=1,\dots,L$. The fermion creation and annihilation operators are
$(C^\dagger_{\mathbf I\alpha}, C_{\mathbf I\alpha}),~\alpha=1,\dots,N_B$.

Consider a complete orthonormal set of functions on the lattice, $\phi^{\mathbf
k}_I$. We call $\mathbf k$, the spectral parameters. When $\phi^{\mathbf k}$
are plane waves, $\phi^{\mathbf k}_I=\frac{1}{L^{d/2}}e^{i\mathbf k\cdot\mathbf
I}$, the spectral parameters are the quasi-momenta taking values in the
Brillouin zone. We define,
\begin{equation}
\label{ccdagkdef}
C_{\mathbf I\alpha}\equiv
\sum_{\mathbf \mathbf k}\phi^{\mathbf k}_{\mathbf I}C_{\mathbf k\alpha},~~~~~
C_{\mathbf k\alpha}=
\sum_{\mathbf I}\left(\phi^{\mathbf k}_{\mathbf I}\right)^*
C_{\mathbf I\alpha}.
\end{equation}

Now consider a mean field state constructed from the single-particle
wavefunctions defined in Section \ref{dgprev}. We consider the case of the
$n^{\rm th}$ band completely filled and all others completely empty
\begin{equation}
\label{mfsdef}
\vert n\rangle\equiv\prod_{\mathbf k,\alpha}\left(u^n_\alpha(\mathbf k)
C^\dagger_{\mathbf k\alpha}\right)\vert 0\rangle
\end{equation}
where $C_{\mathbf k\alpha}\vert 0\rangle=0$. The second order Bargmann
invariant can be written as,
\begin{eqnarray}
\nonumber
{\cal B}^2(\mathbf k_1,\mathbf k_2)&=&
\left(u^{n\dagger}(\mathbf k_1)u^n(\mathbf k_2)\right)
\left(u^{n\dagger}(\mathbf k_2)u^n(\mathbf k_1)\right)\\
\nonumber
&=&-\langle 0\vert
\left(u^{n\dagger}(\mathbf k_1)C_{\mathbf k_1}\right)
\left(u^{n\dagger}(\mathbf k_2)C_{\mathbf k_2}\right)\\
&&\left(C^\dagger_{\mathbf k_1}u^n(\mathbf k_2)\right)
\left(C^\dagger_{\mathbf k_2}u^n(\mathbf k_1)\right)\vert 0\rangle.
\end{eqnarray}
If we can construct a unitary operator, $E(\mathbf k_1,\mathbf k_2)$,
which we call the exchange operator, such that,
\begin{eqnarray}
E(\mathbf k_1,\mathbf k_2)\vert 0\rangle&=&\vert 0\rangle\\
E(\mathbf k_1,\mathbf k_2)
C^\dagger_{\mathbf k_1,\alpha}C^\dagger_{\mathbf k_2\beta}
E^\dagger(\mathbf k_1,\mathbf k_2)&=&
-C^\dagger_{\mathbf k_2,\alpha}C^\dagger_{\mathbf k_1\beta}.
\end{eqnarray}
Then, ${\cal B}^2(\mathbf k_1,\mathbf k_2)$ can be written as the 
expectation value of $E(\mathbf k_1,\mathbf k_2)$ in the two particle
state, $\left(C^\dagger_{\mathbf k_2}u^n(\mathbf k_2)\right)
\left(C^\dagger_{\mathbf k_1}u^n(\mathbf k_1)\right)\vert 0\rangle$.

It is clear that if the exchange operator $E(\mathbf k_1,\mathbf k_2)$ 
commute with all the other fermion creation and annihilation operators, i.e.  
$(C_{\mathbf k\alpha},C^\dagger_{\mathbf k\alpha}),~\mathbf k\ne \mathbf k_1, 
\mathbf k_2$, then,
\begin{equation}
{\cal B}^2(\mathbf k_1,\mathbf k_2)=\langle n\vert E(\mathbf k_1,\mathbf k_2)
\vert n\rangle.
\end{equation}
The exchange operator can be explicitly constructed,
\begin{equation}
\label{e2def}
E(\mathbf k_1,\mathbf k_2)\equiv 
e^{\frac{\pi}{2}\sum_{\alpha=1}^{N_B}
\left(C^\dagger_{\mathbf k_1\alpha}C_{\mathbf k_2\alpha}-h.c\right)}.
\end{equation}

Now consider the third order Bargmann invariant,
\begin{eqnarray}
\nonumber
{\cal B}^3(\mathbf k_1,\mathbf k_2,\mathbf k_3)&=&
\left(u^{n\dagger}(\mathbf k_1)u^n(\mathbf k_3)\right)
\left(u^{n\dagger}(\mathbf k_3)u^n(\mathbf k_2)\right)\\
\nonumber
&&\left(u^{n\dagger}(\mathbf k_2)u^n(\mathbf k_1)\right)\\
\nonumber
&=&\langle 0\vert
\left(u^{n\dagger}(\mathbf k_1)C_{\mathbf k_1}\right)
\left(u^{n\dagger}(\mathbf k_2)C_{\mathbf k_2}\right)\\
\nonumber
&&\left(u^{n\dagger}(\mathbf k_3)C_{\mathbf k_3}\right)
\left(C^\dagger_{\mathbf k_1}u^n(\mathbf k_3)\right)\\
&&\left(C^\dagger_{\mathbf k_3}u^n(\mathbf k_2)\right)
\left(C^\dagger_{\mathbf k_2}u^n(\mathbf k_1)\right) \vert 0 \rangle.
\end{eqnarray}
Again, if we can construct a unitary operator 
${\cal C}(\mathbf k_1,\mathbf k_2,\mathbf k_3)$, such that,
\begin{eqnarray}
\nonumber
{\cal C}(\mathbf k_1,\mathbf k_2,\mathbf k_3)
C^\dagger_{\mathbf k_3\alpha}
C^\dagger_{\mathbf k_2\beta}
C^\dagger_{\mathbf k_1\gamma}
{\cal C}^\dagger(\mathbf k_1,\mathbf k_2,\mathbf k_3)
&=&C^\dagger_{\mathbf k_1\alpha}C^\dagger_{\mathbf k_3\beta}
C^\dagger_{\mathbf k_2\gamma}\\
{\cal C}(\mathbf k_1,\mathbf k_2,\mathbf k_3)\vert 0\rangle&=&\vert 0\rangle
\end{eqnarray}
${\cal C}$ can be constructed in terms of the exchange operators,
\begin{equation}
\label{c3def}
{\cal C}(\mathbf k_1,\mathbf k_2,\mathbf k_3)=
E(\mathbf k_1,\mathbf k_3)E(\mathbf k_3,\mathbf k_2).
\end{equation}

It is clear that the procedure will generalise the higher order Bargmann
invariants as well. Thus we have shown that the Bargmann invariants, 
defined in terms of the single-particle states, can be expressed as the 
expectation values of the exchange operators and cyclic operators constructed
from them in mean field states. We have shown this above for the mean field 
states with only one band filled. However, it is not difficult to generalize
it for an arbitrary number of filled bands. 

\subsection{Quantum distances for correlated states}

The results of the above section motivates us to define quantum distance
between two points in the spectral parameter space in terms of the 
expectation values of the exchange  operators. 

To do so we define the Fock basis for the 
many-body states. We denote the occupation numbers of the 
$(\mathbf k,\alpha)$ mode by $n_{\mathbf k,\alpha}$.
The collection of all the occupation numbers is denoted by $\left\{n\right\}$.
The empty state ($n_{\mathbf k,\alpha}=0,~\forall\mathbf k,\alpha$) is denoted
by $\vert 0\rangle$. The Fock basis is,
\begin{equation}
\label{fbdef}
\vert\left\{ n\right\}\rangle=\prod_{\mathbf k,\alpha}
\left(C^\dagger_{\mathbf k\alpha}\right)^{n_{\mathbf k\alpha}}\vert 0\rangle,~~
C^\dagger_{\mathbf k\alpha}C_{\mathbf k\alpha}\vert\left\{n\right\}\rangle
=n_{\mathbf k\alpha}\vert\left\{n\right\}\rangle.
\end{equation}
We also need to define the ordering of the operators in Eq.~\eqref{fbdef}.
Since the set $(\mathbf k,\alpha)$ is countable, we associate a unique integer
$m(\mathbf k,\alpha)$ with every element of the set. With the definitions,
\begin{equation}
C_m\equiv C_{\mathbf k\alpha},~~n_m\equiv n_{\mathbf k\alpha}.
\end{equation}
The ordering is defined as,
\begin{equation}
\label{fborder}
\vert\left\{ n\right\}\rangle=\prod_{m=1}^{N_BL^d}
C^\dagger_m\vert 0\rangle.
\end{equation}
Any many-body state, $\vert\psi\rangle$ can be expanded as,
\begin{equation}
\label{psiexp}
\vert\psi\rangle=\sum_{\{n\}}\psi(\{n\})\vert\{n\}\rangle.
\end{equation}
We define the exchange operators, $E(\mathbf k_1,\mathbf k_2)$, by their action
on the Fock basis. These operators exchange the occupation numbers of the modes
at $\mathbf k_1$ and $\mathbf k_2$. We first define,
\begin{equation}
\label{ek1k2alphadef}
E_\alpha(\mathbf k_1,\mathbf k_2)\vert \dots,n_{\mathbf k_1\alpha},\dots,
n_{\mathbf k_2\alpha},\dots\rangle\equiv
\vert ..,n_{\mathbf k_2\alpha},..,n_{\mathbf k_1\alpha},..\rangle.
\end{equation}
The exchange operator is then defined as,
\begin{equation}
\label{ek1k2def}
E(\mathbf k_1,\mathbf k_2)\equiv\prod_{\alpha=1}^{N_B}
E_\alpha(\mathbf k_1,\mathbf k_2).
\end{equation}
From Eqs.~\eqref{ek1k2alphadef} and \eqref{ek1k2def}, it follows that,
\begin{eqnarray}
\nonumber
E(\mathbf k,\mathbf k)=I&,&
E(\mathbf k_1,\mathbf k_2)=E(\mathbf k_2,\mathbf k_1)\\
\nonumber
E^\dagger(\mathbf k_1,\mathbf k_2)=E^{-1}(\mathbf k_1,\mathbf k_2)&,&
E^\dagger(\mathbf k_1,\mathbf k_2)=E(\mathbf k_1,\mathbf k_2)\\
\label{eprop}
E^2(\mathbf k_1,\mathbf k_2)=I.
\end{eqnarray}

The second order Bargmann invariants, for a general many-particle
state $\vert\psi\rangle$ is defined as,
\begin{equation}
\label{b2csdef}
{\cal B}^2(\mathbf k_1,\mathbf k_2)\equiv
\langle\psi\vert E(\mathbf k_1,\mathbf k_2)\vert\psi\rangle.
\end{equation}
The quantum distance between $\mathbf k_1$ and $\mathbf k_2$ is then
defined as in Eq.~\eqref{dpsi1psi2def},
\begin{equation}
\label{dk1k2def}
d(\mathbf k_1,\mathbf k_2)\equiv\sqrt{1
-\left({\cal B}^2(\mathbf k_1,\mathbf k_2)\right)^\alpha}.
\end{equation}

\subsection{The triangle inequalities}
\label{tineq}

We define the quantum distance between $\mathbf k_1$ and $\mathbf k_2$ as the 
quantum distance between the states $\vert\psi\rangle$
and $\vert\chi(\mathbf k_1,\mathbf k_2)\rangle=
E(\mathbf k_1,\mathbf k_2)\vert\psi\rangle$. We
denote it by,
\begin{equation}
\label{capddef}
D^2(\psi,\chi(\mathbf k_1,\mathbf k_2))=
1-
\left\vert\langle\chi(\mathbf k_1,\mathbf k_2)\vert\psi\rangle\right\vert^\alpha.
\end{equation}
Our definition of the distance between $\mathbf k_1$ and $\mathbf k_2$ is,
\begin{equation}
\label{dcapd}
d(\mathbf k_1,\mathbf k_2)=D(\psi,\chi(\mathbf k_1,\mathbf k_2)).
\end{equation}

We use the Ptolemy inequality, which holds in any Hilbert
space \cite{Ptolemy}, to prove that our definition of the distance in Eq.~\ref{dcapd}
satisfies the triangle inequality. For $\alpha=2$, the problem reduces to 
the classical problem in Euclidean space with the standard definition of
distance. We show this in appendix \ref{2euc}.

The Ptolemy inequality states
that six distances between any four points, $d_{ij},~i,j=1,\dots,4$
satisfies the following inequalities
\begin{equation}
\label{ptolemy}
d_{ij}d_{kl}+d_{ik}d_{jl}\ge d_{il}d_{jl}
\end{equation}
where $(i,j,k,l)$ are distinct.

Consider any three points in the spectral parameter space, $\mathbf k_1,
~\mathbf k_2$ and $\mathbf k_3$. Define,
\begin{eqnarray}
\nonumber
\vert\chi_1\rangle\equiv E(\mathbf k_2,\mathbf k_3)\vert\psi\rangle&,&
\vert\chi_2\rangle\equiv E(\mathbf k_3,\mathbf k_1)\vert\psi\rangle,\\
\label{chi123def}
\vert\chi_3\rangle\equiv E(\mathbf k_1,\mathbf k_2)\vert\psi\rangle.
\end{eqnarray}
The Ptolemy inequality implies,
\begin{align}
\label{capdptineq}
D(\psi,\chi_1)D(\chi_2,\chi_3)+D(\psi,\chi_2)D(\chi_3,\chi_1)\ge \nonumber \\
D(\psi,\chi_3)D(\chi_1,\chi_2).
\end{align}
We will now show that $D(\chi_1,\chi_2)=D(\chi_2,\chi_3)=D(\chi_3,\chi_1)$.
From the definitions in Eq.~\eqref{chi123def} and Eq.~\eqref{eprop},
\begin{eqnarray}
\label{chioverlaps1}
\langle\chi_1\vert\chi_2\rangle&=&\langle\psi\vert E(\mathbf k_2,\mathbf k_3)
E(\mathbf k_3,\mathbf k_1)\vert\psi\rangle\\
\label{chioverlaps2}
\langle\chi_2\vert\chi_3\rangle&=&\langle\psi\vert E(\mathbf k_3,\mathbf k_1)
E(\mathbf k_1,\mathbf k_2)\vert\psi\rangle\\
\label{chioverlaps3}
\langle\chi_3\vert\chi_1\rangle&=&\langle\psi\vert E(\mathbf k_1,\mathbf k_2)
E(\mathbf k_2,\mathbf k_3)\vert\psi\rangle.
\end{eqnarray}
From the definition of the exchange operators in Eq.~\eqref{ek1k2alphadef},
it is easy to show that
\begin{eqnarray}
\nonumber
E(\mathbf k_2,\mathbf k_3)E(\mathbf k_3,\mathbf k_1)
&=&E(\mathbf k_3,\mathbf k_1)E(\mathbf k_1,\mathbf k_2)\\
&=&E(\mathbf k_1,\mathbf k_2)E(\mathbf k_2,\mathbf k_3).
\end{eqnarray}
All the above three operators cyclically permute the occupation numbers
as $(n_{\mathbf k_1\alpha},n_{\mathbf k_2\alpha},n_{\mathbf k_3\alpha})
\rightarrow (n_{\mathbf k_2\alpha},n_{\mathbf k_3\alpha},n_{\mathbf k_1\alpha})$
. Hence we have,
\begin{equation}
\label{dchieq}
D(\chi_1,\chi_2)=D(\chi_2,\chi_3)=D(\chi_3,\chi_1).
\end{equation}
Hence, if $D(\chi_1,\chi_2)\ne 0$, Eqs.~\eqref{capdptineq}
and \eqref{dchieq} imply
\begin{equation}
\label{capdtineq}
D(\psi,\chi_1)+D(\psi,\chi_2)\ge D(\psi,\chi_3).
\end{equation}
If $D(\chi_1,\chi_2)=0$, then it implies that the three states 
$\vert\chi_i\rangle,~i=1,2,3$ are the same up to overall phases. 
We then have $D(\psi,\chi_1)=D(\psi,\chi_2)=D(\psi,\chi_3)$ so
that above inequality is still satisfied.

Eqs.~\eqref{capdtineq} and \eqref{dcapd} imply,
\begin{equation}
\label{dtineq}
d(\mathbf k_2,\mathbf k_3)+d(\mathbf k_3,\mathbf k_1)
\ge d(\mathbf k_1,\mathbf k_2).
\end{equation}
The definition of the distance in Eq.~\eqref{dcapd} hence 
satisfies the triangle inequalities.

\subsection{Geometric phases}

It is natural to attempt to define the geometric phases associated with
loops in the spectral parameter space as the expectation value of the
loop operator constructed in Eq.~\eqref{c3def}. However, it has to
satisfy the additive laws in Eqs.~\eqref{addlaw1},\eqref{addlaw2}.
We have checked this numerically for random states and find that there
are a large number of violations. So while the definition reproduces the
single-particle results for mean field states, it is not a meaningful
generalization for correlated states.

We leave this issue of defining geometric phases associated with loops in
the spectral parameter space for correlated states for future work and 
concentrate on the quantum distances for the rest of this paper.

\section{The exchange operators in terms of the fermion operators}
\label{efops}

In this section, we construct the exchange operators, for the general
many-particle state, in terms of the fermion operators.

We define unitary operators,
\begin{eqnarray}
\label{udef}
U_\alpha(\mathbf k_1,\mathbf k_2)&\equiv&
e^{i\frac{\pi}{2}T_\alpha(\mathbf k_1,\mathbf k_2)},\\
\label{tdef}
T_\alpha(\mathbf k_1,\mathbf k_2)&\equiv&
\frac{1}{i}\left(C^\dagger_{\mathbf k_1\alpha}C_{\mathbf k_2\alpha}-
C^\dagger_{\mathbf k_2\alpha}C_{\mathbf k_1\alpha}\right).
\end{eqnarray}

It is easily shown that,
\begin{eqnarray}
\nonumber
U^\dagger_\alpha(\mathbf k_1,\mathbf k_2)
\left(C^\dagger_{\mathbf k_1\alpha}\right)^{n_{\mathbf k_1\alpha}}
U_\alpha(\mathbf k_1,\mathbf k_2)
&=&\left(C^\dagger_{\mathbf k_2\alpha}\right)^{n_{\mathbf k_1\alpha}}\\
\nonumber
U^\dagger_\alpha(\mathbf k_1,\mathbf k_2)
\left(C^\dagger_{\mathbf k_2\alpha}\right)^{n_{\mathbf k_2\alpha}}
U_\alpha(\mathbf k_1,\mathbf k_2)
&=&(-1)^{n_{\mathbf k_2\alpha}}
\left(C^\dagger_{\mathbf k_1\alpha}\right)^{n_{\mathbf k_2\alpha}}.\\
\label{ucnd}
\end{eqnarray}

We compute the action of $U^\dagger$ on the two particle states to be,
\begin{eqnarray}
\nonumber
U^\dagger_\alpha(\mathbf k_1,\mathbf k_2)
\vert n_{\mathbf k_1\alpha},n_{\mathbf k_2\alpha}\rangle
&=&(-1)^{n_{\mathbf k_2\alpha}(1-n_{\mathbf k_1\alpha})}
\vert n_{\mathbf k_2\alpha},n_{\mathbf k_1\alpha}\rangle.\\
\label{un1n2}
\end{eqnarray}
Thus if we define,
\begin{equation}
\label{ec2def}
\tilde E_\alpha(\mathbf k_1,\mathbf k_2)\equiv
e^{i\pi\left(\rho_\alpha(\mathbf k_1)(1-\rho_\alpha(\mathbf k_2))\right)}
U^\dagger_\alpha(\mathbf k_1,\mathbf k_2),
\end{equation}
then we have,
\begin{equation}
\tilde E_\alpha(\mathbf k_1,\mathbf k_2)
\vert n_{\mathbf k_1\alpha},n_{\mathbf k_2\alpha}\rangle=
\vert n_{\mathbf k_2\alpha},n_{\mathbf k_1\alpha}\rangle.
\end{equation}
This is the desired action of the exchange operator on the 
two particle states. For the general Fock space basis state,
we have to take into account the ordering of the fermion operators
in the definition of the basis states. As defined in Eq.~\eqref{fborder},
we order the spectral parameters by associating a natural number,
$m(\mathbf k,\alpha)$ with each mode. Using the fact that 
$\left(C^\dagger_m\right)^{n_m}\left(C^\dagger_l\right)^{n_l}=
(-1)^{n_mn_l}\left(C^\dagger_l\right)^{n_l}\left(C^\dagger_m\right)^{n_m}$,
we define,
\begin{equation}
\label{ecgendef}
E_\alpha(\mathbf k_1,\mathbf k_2)\equiv
e^{i\pi\nu_\alpha(\mathbf k_1,\mathbf k_2)}
\tilde E_\alpha(\mathbf k_1,\mathbf k_2)
\end{equation}
where $\nu_\alpha(\mathbf k_1,\mathbf k_2)$ has the following form,
\begin{equation}
\label{nudef}
\nu_\alpha(\mathbf k_1,\mathbf k_2)\equiv
(\rho_\alpha(\mathbf k_1)+\rho_\alpha(\mathbf k_2))
\sum_{l=m(\mathbf k_1,\alpha)+1}^{m(\mathbf k_2,\alpha)-1} 
\rho_\alpha(\mathbf k_l).
\end{equation}
The exchange operator defined above in Eq.~\eqref{ecgendef} has the 
desired action on the Fock basis.

\section{Application to the 1-dimensional $t-V$ model}
\label{tv1}

In this section, we apply our formalism to explore the geometry of the 
ground state of the one dimensional $t-V$ model. This is a simple but 
non-trivial correlated state. The model is time-reversal
and parity invariant. So we do not expect any geometric phase effects and
hence concentrate on the quantum distances.

The Hamiltonian is,
\begin{equation}
\label{tvham}
H=\sum_{i=1}^L\left(-t\left(C^\dagger_{i+1}C_i+h.c\right)+Vn_in_{i+1}\right)
\end{equation}
where $C^\dagger_i,~C_i$ are the fermion creation and annihilation operators.
$n_i\equiv C^\dagger_iC_i$ is the operator representing the fermion number
density at the $i^{\rm th}$ site. We will concentrate on the half-filled
states, namely $\sum_i n_i\vert\psi\rangle=L/2\vert\psi\rangle$.

This is a well studied model \cite{Yang-XXZ,Baxter,tV-bosonisation}. At $V=0$, 
the ground state is a simple non-interacting Fermi Sea (FS).
As soon as the interaction is turned on,the ground state is a metallic Luttinger liquid.  
The long-distance correlations of this state decay as power laws, characterised by an anomalous
dimension of the fermion operators \cite{Luttinger,Haldane_LL_CE,Haldane_LL}. 
The anomalous dimension varies continuously as a function of $V$ . 
At $V=2$, there is a transition to an insulating charge density wave state (CDW) \cite{Shankar}, 
in which the translational symmetry is spontaneously broken.

We apply our formalism to compute the quantum distances in this model and 
explore how the corresponding geometry reflects the physics described above.
In this work, we choose the spectral parameter to be the quasi-momenta. The
Fourier transform of the fermion operators are defined as,
\begin{equation}
\label{ftdef}
C_k=\frac{1}{\sqrt L}\sum_{i=1}^L~e^{-i\frac{2\pi}{L}ki}C_i,
\end{equation}
where $k$ is an integer that we choose to be $-L/2\le k< L/2$.

In this one band model, for translationally invariant states, Eq.~\eqref{ecgendef}
can be used to derive the following expression for the expectation values of the exchange operators,
\begin{equation}
\label{en1n2}
\langle E(k_1,k_2)\rangle=1
-\left\langle\left(C^\dagger_{k_1}C_{k_1}-C^\dagger_{k_2}C_{k_2}\right)^2
\right\rangle.
\end{equation}

The ground states for the extreme limits of the interaction strength ($V=0$ 
and $V=\infty$) are simple and the exact distance matrices can be obtained 
analytically. We present the solutions below.

\section{The exact distance matrices at $V=0,\infty$}
\label{edm}

\subsection{$V=0$}

The Hamiltonian (\ref{tvham}), at $V=0$ is,
\begin{equation}
\label{t0ham}
H^0=-\sum_{k=-L/2}^{L/2-1}~2t\cos\left(\frac{2\pi}{L}k\right)C^\dagger_kC_k
\end{equation}
For convenience, we choose $L$ to be an even number which is not divisible by
$4$. Further we define $N=L/2$ (an odd number). The ground state of the above 
Hamiltonian, for a system with $N$ fermions is,
\begin{equation}
\label{fsdef}
\vert FS\rangle=\prod_{k=-N/2-1/2}^{N/2+1/2}C^\dagger_k\vert 0\rangle
\end{equation}
where $\vert 0\rangle$ is the empty state, defined by $C_i\vert 0\rangle=0,~
\forall i$.

It is easy to compute the expectation value of the exchange operator. 
If we exchange the occupation numbers of two quasi-momenta which are both 
in the Fermi sea or both outside it, the physical state is unchanged. 
Hence the expectation value of the exchange operator is $\pm 1$. On the 
other hand when one quasi-momenta is in the Fermi sea and the other outside it, the exchange operator 
removes a particle from the Fermi sea and creates one outside is. This particle-hole state is orthogonal 
to $\vert FS\rangle$ and hence the expectation value of the exchange operator is 0.
\begin{eqnarray}
\nonumber
\vert\langle E(n,m)\rangle_{FS}\vert=1 ~~&(&n\in FS,~m\in FS)\\
~{\rm or}~&(&n\notin FS,~m\notin FS)\\
\nonumber
           =0~~&(&n\in FS,~m\notin FS)\\
           ~{\rm or}~&(&n\notin FS,~m\in FS).
\end{eqnarray}
The squared distances are thus,
\begin{eqnarray}
\nonumber
\left(d^{FS}(n,m)\right)^2=&1&-\vert\langle E(n,m)\rangle\vert^\alpha\\
\nonumber
=&0&~(n\in FS~{\rm and}~m\in FS)\\
&~{\rm or}&~(n\notin FS,~m\notin FS)\\
\nonumber
=&1&~(n\in FS~{\rm and}~m\notin FS)\\
&~{\rm or}&~(n\notin FS,~m\in FS).
\end{eqnarray}
To write the distance matrix in a compact form, it is convenient to relabel
the momenta as $n\rightarrow n,p$ with $-N/2\le n<N/2,~p=\pm$ as follows,
\begin{eqnarray}
n-&=&n,~~~-N/2\le n<N/2\\
n+&=&n+N~~-N/2\le n<N/2.
\end{eqnarray}
Thus $n-\in FS$ and $n+\notin FS$. We define an $N\times N$ matrix, 
${\cal I}$ with all entries equal to 1, ${\cal I}_{nm}=1$
and a $2\times2$ matrix, $\tau^x_{pp'}=1-\delta_{pp'}$.
The distance matrix can then be written as,
\begin{equation}
\label{dfsmat1}
d^{FS}(np,mp')={\cal I}_{nm}\tau^x_{pp'}~\Rightarrow~
d^{FS}=\left(\begin{array}{cc}0&{\cal I}\\{\cal I}&0\end{array}\right).
\end{equation}

\subsection{$V=\infty$}

At $V=\infty$, the Hamiltonian (\ref{tvham}) is,
\begin{equation}
\label{0vham}
H_\infty=V\sum_i n_in_{i+1}.
\end{equation}
In the thermodynamic limit, translation symmetry is spontaneously broken and
there are two degenerate ground states. One in which all the particles are 
localised at the even sites and the other where all are localised at the 
odd sites. For finite $L$ (and very small $t$) the degeneracy splits and 
the symmetric combination is the ground state. We denote it by 
$\vert CDW\rangle$
\begin{equation}
\label{cdwdef}
\vert CDW\rangle\equiv\frac{1}{\sqrt 2}\left(\prod_iC^\dagger_{2i}\vert 0\rangle
+\prod_iC^\dagger_{2i+1}\vert 0\rangle\right).
\end{equation}
The expectation values of the exchange operators are,
\begin{eqnarray}
\nonumber
\vert\langle E(n,m)\rangle_{CDW}\vert&=&1~~n=m\\
\nonumber
                                     &=&0~~n=m+\frac{L}{2}\\
\nonumber
                                     &=&\frac{1}{2},~n\ne m,~n\ne m+\frac{L}{2}.
\end{eqnarray} 
Consequently, the distance matrix elements are,
\begin{eqnarray}
\nonumber
d^{CDW}(n,m)&=&0,~~~~~~~~~~~~n=m\\
            &=&1,~~~~~~~~~~~~n=m+\frac{L}{2}\\
            &=&\sqrt{1-\frac{1}{2^\alpha}},~n\ne m,~n\ne m+\frac{L}{2}.
\end{eqnarray}
We define $c(\alpha)\equiv\sqrt{1-1/2^\alpha}$ and denote the $N\times N$
identity matrix by $I$, $I_{nm}=\delta_{nm}$. The distance matrix can then
be written as,
\begin{equation}
\label{dmatcdw}
d^{CDW}=c(\alpha)\left(\begin{array}{cc}{\cal I}-I&{\cal I}-I\\
{\cal I}-I&{\cal I}-I\end{array}\right)
+\left(\begin{array}{cc} 0&I\\I&0\end{array}\right).
\end{equation}

\subsection{Discussion}
\label{edd}

The most striking aspect of the distance matrices in the two extreme
limits is that whereas the points are highly ``clustered" at $V=0$,
they are completely ``spread out" in the $V=\infty$ case. More precisely,
the distance matrix at $V=0$ is the same as a space with only two points
with distance 1 between them. All the quasi-momenta in the Fermi sea map
on to one point and all the points outside it map on to the other point.
However, the distance matrix at $V=\infty$ seems to correspond to a space
with a thermodynamic number of dimensions. In particular, if we model the 
distances as those between points in a Euclidean space \cite{DG}, then it turns 
out that these points lie on a $L-1$ dimensional sphere.

In particular, at $V=0$, the distance matrix reveals a sharp Fermi surface, in
the sense that the distances are discontinuous across it, whereas at $V=\infty$
there is no signal of it. All the points in the latter case look identical.

Another manifestation of the same feature is that, in the $V=\infty$ case, if
we pick any three quasi-momenta, the distances between them either correspond
to an equilateral triangle or an isosceles triangle (when $k$ and $k+\pi$ are
two of the three quasi-momenta). On the other hand at $V=0$ the distances
between any three quasi-momenta corresponds to a single point or a segment.

In the next section, we examine these three aspects for finite, non-zero 
values of the interaction.

\section{Numerical distance matrices at $0<V<\infty$}
\label{ndm}

We numerically diagonalise the Hamiltonian, in the quasi-momentum basis,
for the 18-site system, for values of interaction strength $V=0-12$. 
Since we obtain the numerical ground state in the quasi-momentum occupation number basis, 
it is easy to act the exchange operators on it and hence compute the quantum distance matrix. 
All the computations reported in this paper are done at $\alpha=2$. We describe our
results below.

\subsection{Overall structure of the distance matrix}

\begin{figure}
\includegraphics[width=0.21\textwidth]{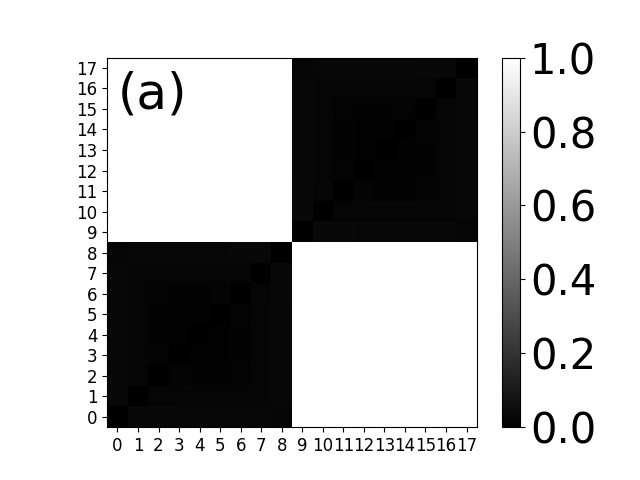}
\includegraphics[width=0.21\textwidth]{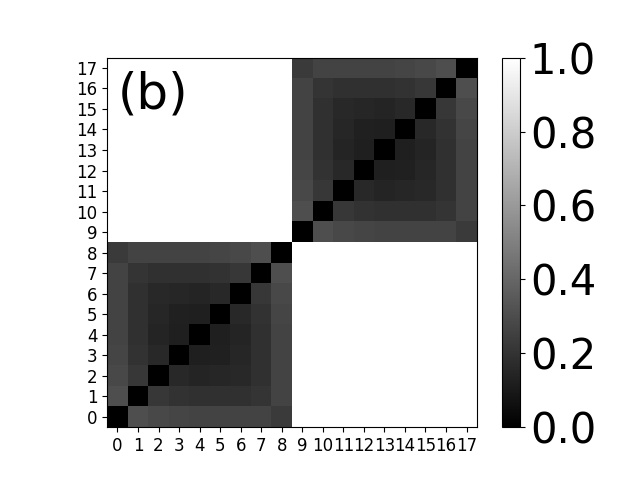}\\
\includegraphics[width=0.21\textwidth]{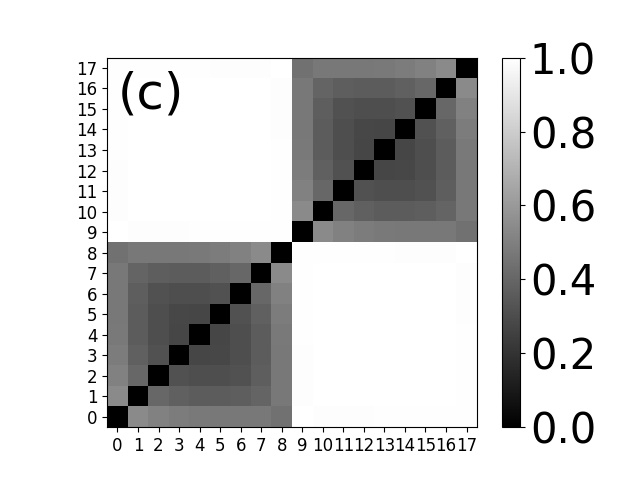}
\includegraphics[width=0.21\textwidth]{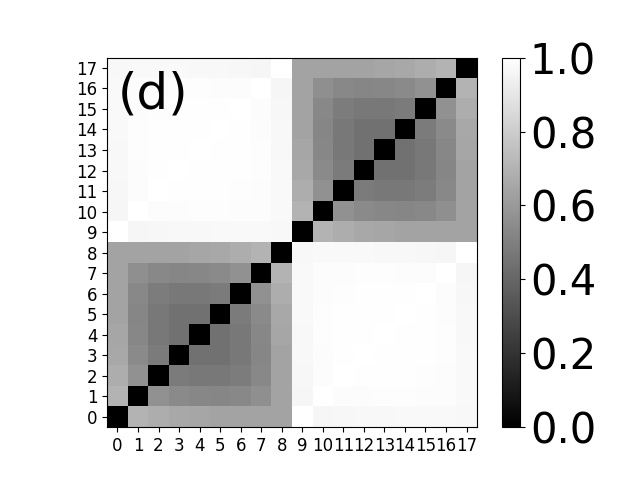}\\
\includegraphics[width=0.21\textwidth]{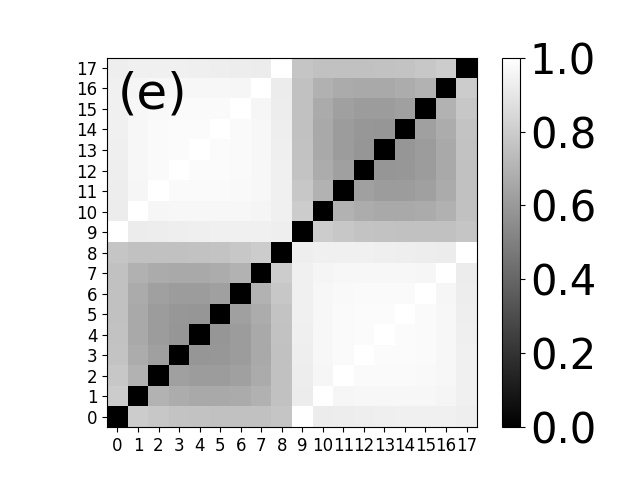}
\includegraphics[width=0.21\textwidth]{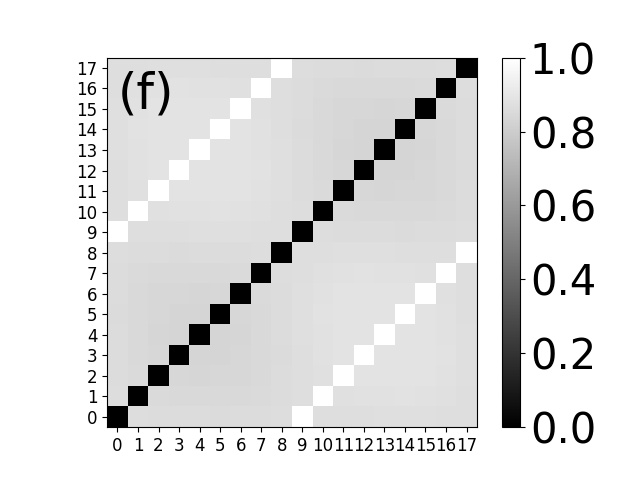}\\
\caption{(a)-(f) Distance matrices obtained from numerical computation for interaction strengths $V=0.1$ (a), $V=1$ (b), $V=2$ (c), $V=3$ (d), $V=4$ (e) and $V=12$ (f).}
\label{Matrix}
\end{figure}

As soon as we turn the interaction on, the distances between pairs of
quasi-momenta in the Fermi sea (and pairs outside it) are no longer zero
and are not all equal either. Also, the distances between quasi-momenta
in the Fermi sea and outside it is no longer equal and also not equal to 1.
The distance matrix is of the form,
\begin{equation}
d=\left(\begin{array}{cc}\Delta&\Delta_e\\\Delta_e&\Delta\end{array}\right)
\end{equation}
where, $\Delta$ has all matrix elements $<< 1$ and $\Delta_e$ has matrix elements
slighlty less than 1. As the interaction strength increases, the matrix 
elements of $\Delta$ increase and those of $\Delta_e$ decrease. By 
$V\approx 4$, the features of the matrix characterising $V=\infty$\hspace{0.02cm} limit start manifesting. The 
evolution of the matrix is shown in Fig.~(\ref{Matrix}).

\subsection{Distances from $k=-\pi$}

\begin{figure}
\includegraphics[angle=0,width=0.45\textwidth]{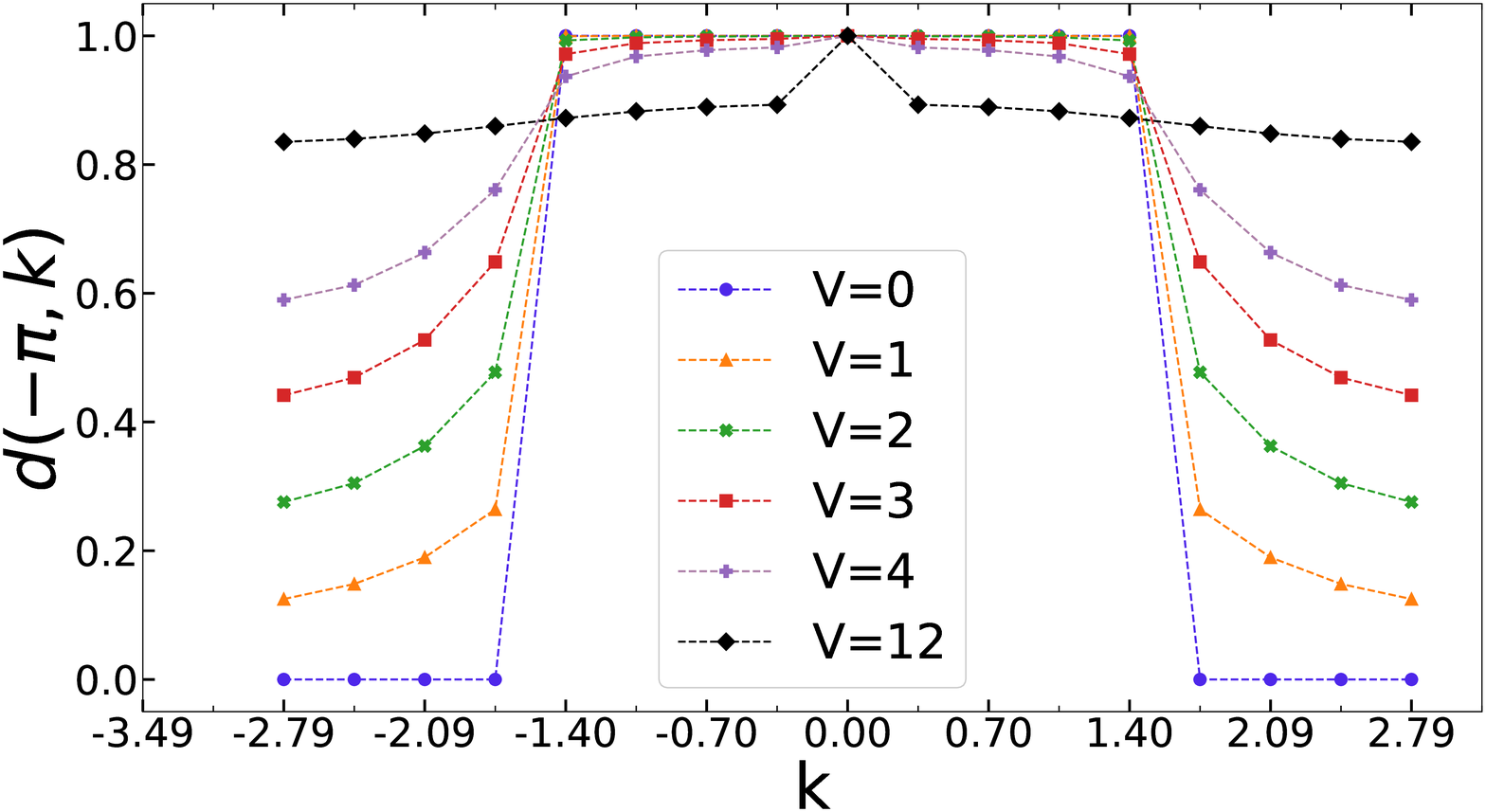}
\caption{Distance $d(-\pi,k)$ between $k=-\pi$ and the other $k$ modes in the Brillouin zone (BZ)
for different values of the interaction strength $V$.}
\label{Pi-fig}
\end{figure}

Figure (\ref{Pi-fig}) shows the distance, $d(-\pi,k)$ for different interaction 
strengths. The Fermi points are $k_F=\pm\pi/2$.
At $V=0$, the distance jumps from 0 to 1 across the Fermi points. At small
$V$, the discontinuity seems to persist and at large $V$, there is
no discontinuity.

\begin{figure}
\centering
\includegraphics[angle=0,width=0.48\textwidth]{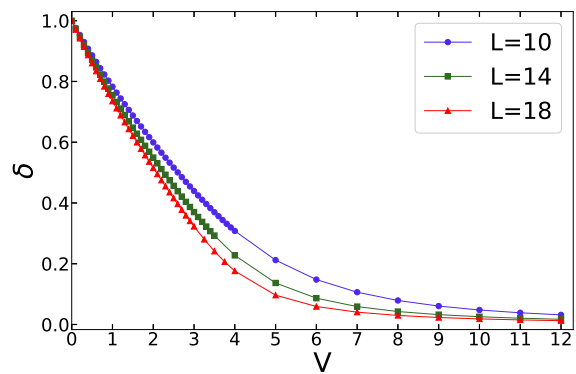}
\caption{$\delta=d(-\pi,-\dfrac{\pi}{2})-d(-\pi,-\dfrac{\pi}{2}-\dfrac{2\pi}{L})$, gives a measure of the discontinuity across the 
Fermi points. It is studied as a function of interaction strength $V$ for different system sizes.}
\label{Pi-system}
\end{figure}

We examine this more closely by plotting $\delta$ which is the difference between $d(-\pi,-\frac{\pi}{2})$ and
$d(-\pi,-\frac{\pi}{2}-\frac{2\pi}{L})$ for different system sizes in Fig.~(\ref{Pi-system}).
The discontinuity is insensitive to the system size for $V\lessapprox 2$ and starts 
depending on the system size for larger values of $V$ indicating that the
discontinuity may persist in the thermodynamic limit at small values of $V$.
However, at $L=18$, we are far from the thermodynamic limit. While it seems
clear that there is a very sharp change across the Fermi point, we cannot
conclusively say if it is a discontinuity. The thermodynamic limit is 
accessible at small $V$ by the bosonisation technique. We are currently 
investigating this and will be reporting it in a future publication.

An interesting feature in Fig.~(\ref{Pi-fig}) is that $d(-\pi,0)=1$ for all values
of the interaction. Indeed we find that $d(k,k+\pi)=1$ for
all values of $k$ and $V$. From Eq.~\eqref{en1n2} it can be inferred
that this is a consequence of the particle-hole symmetry of the model,
$C^\dagger_k\rightarrow C_{k+\pi}$.

\subsection{Nearest neighbour distances}

\begin{figure}
\includegraphics[angle=0,width=0.48\textwidth]{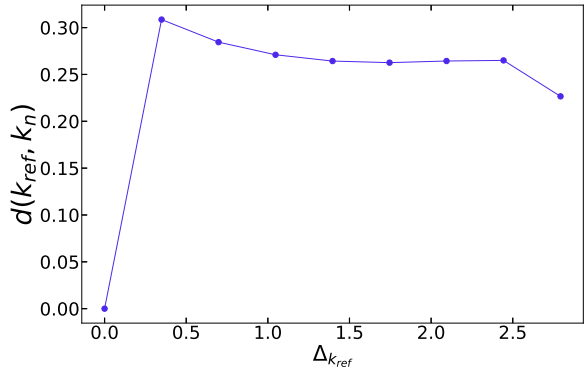}
\caption{Distances between $k_{ref}=-\dfrac{\pi}{2}$ and other modes $k_{n}$, where $k_{n} \in FS$,
as a function of $\Delta_{k_{ref}}$, which is the seperation between them in the BZ  given by 
$\Delta_{k_{ref}}=k_{n}-k_{ref}$, for $V=1$.}
\label{D-func_k}
\end{figure}

We find that the distances between two quasi-momenta do not decrease 
monotonically with the separation between them. The $V=\infty$ case is an 
extreme example, however, as shown in Fig.~(\ref{D-func_k}) this is true even for a small value of
$V$. At $V=1$, we find  for the distances from a reference $k$ mode $-\pi/2$ ($k_{ref}$), 
the distance from the closest $k$ mode is infact having the optimum value.
This indicates that the quantum metric $g(k)$, defined by
$\lim_{\Delta k\rightarrow 0} d^2(k,k+\Delta k)=g(k)\Delta k^2$, may not be
well defined in this system. 

\begin{figure}
\includegraphics[angle=0,width=0.48\textwidth]{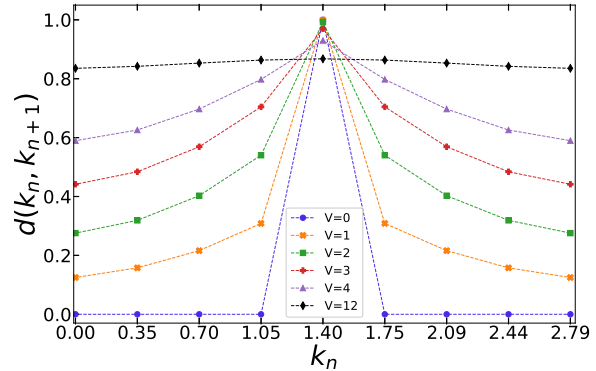}
\caption{Nearest Neighbour Distance for different interaction strength $V$, over half the BZ.}
\label{Dist_nn}
\end{figure}

The nearest neighbour distance $d(k_n,k_{n+1})$ is plotted for different $V$
in Fig.~(\ref{Dist_nn}) over half the Brillioun zone, over the full BZ the value of $n$ 
runs from $0$ ($k_{0} \equiv -\pi$) to $L-1$ ($k_{L-1} \equiv \pi- \frac{2\pi}{L}$) and we consider $k_{L} \equiv -\pi$. 
At $V=0$ there are all zeros except at the Fermi point, when
one quasi-momentum is in the Fermi sea and it's nearest neighbour is outside
it, in which case it is equal to 1. Hence there is a delta function singularity
at the Fermi point. At low $V$ this singularity remains but gets smoothened
out at large $V$. At $V=\infty$ all the nearest neighbour distances are equal
and it can be seen that this is almost the case at $V=12$.

The clustering feature discussed in Section \ref{edd} is nicely illustrated
by the nearest neighbour distances using the following construction which
represents them on a unit circle.

We first define a radius, $R$, in terms of the sum of all the nearest neighbour
distances,
\begin{equation}
\label{cdef}
2\pi R\equiv\sum_{i=0}^{L-1} d(k_i,k_{i+1}).
\end{equation}
This radius varies with $V$, from $R=1/\pi$ at $V=0$ to $R=c(\alpha)L/(2\pi)$ at $V=\infty$.

Each nearest neighbour distance is represented by an angle,
\begin{equation}
\Delta\theta_{i,i+1}=\frac{d(k_i,k_{i+1})}{R}.
\end{equation}

Finally, each quasi-momentum is represented by an angle,
\begin{equation}
\theta_{k_i}=\sum_{j=0}^i\Delta\theta_{j-1,j}
\end{equation}
where, $\Delta\theta_{-1,0}\equiv 0$.

\begin{figure}
\includegraphics[width=0.45\textwidth]{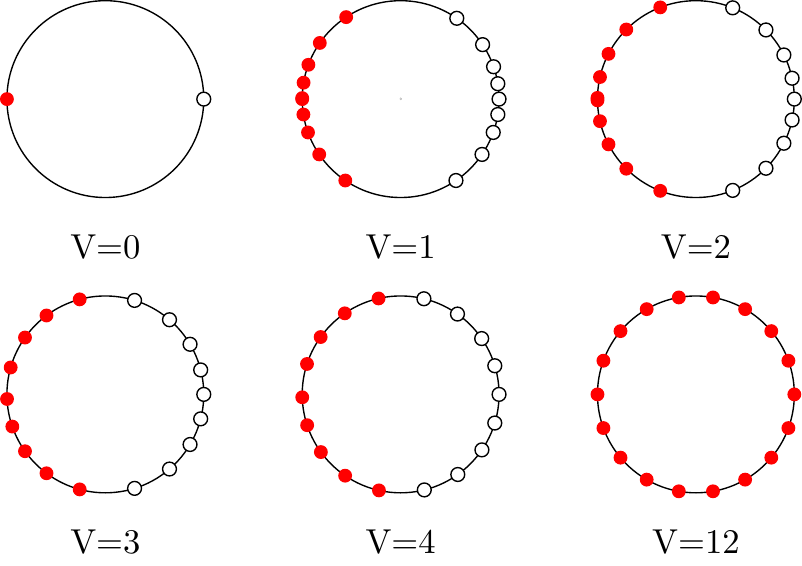}
\caption{Schematic representations on unit circle for different interaction $V$. For each unit circle, in the first five cases ($V=0-4$),
the smaller filled circles on the unit circle represent modes inside the Fermi sea and the smaller open circles correspond 
to modes outside the Fermi sea. For the sixth unit circle ($V=12$) all the modes are equally spaced and represented by smaller filled circles.}
\label{circle-fig}
\end{figure}

The points on the unit circle, as defined above, are plotted in Fig. (\ref{circle-fig}).
At $V=0$ all the points collapse into $\theta=0,\pi$,
at small $V$, they spread out but the points in the Fermi sea and those outside
it are well seperated. At $V$ between 2 and 3, the seperation starts closing
and at $V\ge 4$ the seperation is almost indistinguishable from the $V=\infty$
case when they are equally spaced.

\subsection{Structure of triangles}

We now consider the structure of the triangles corresponding to the 
three distances between three quasi-momenta. At $V=\infty$, most of the 
triangles are equilateral triangles (except those that contain $k$ and
$k+\pi$). We consider only the equilateral triangles. At $V=0$, as mentioned 
earlier, there are no triangles, only points and segments. 

The triangles are of two types, one formed of all three quasi-momenta in
the Fermi sea (or all three outside it). We refer to these as particle 
triangles. The other type are those formed by two quasi-momenta in the Fermi
sea and one outside (or the other way around). We refer to these as 
particle-hole triangles.

\begin{figure}
\includegraphics[width=0.45\textwidth]{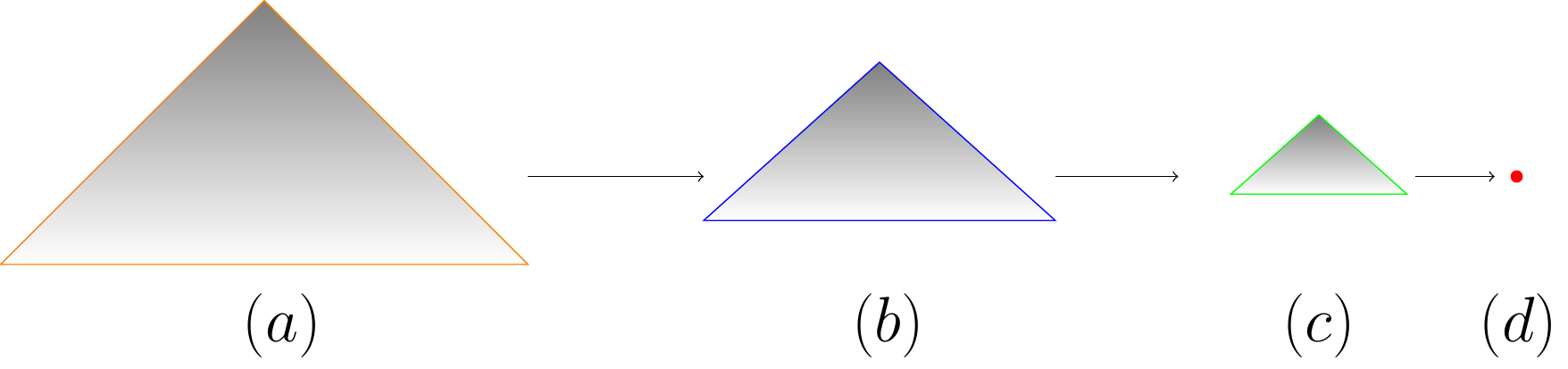}
\caption{(Color online) $(a)-(c)$ Particle triangles for values of interaction strength $V=4$ (a) (orange), 
$V=2$ (b) (blue) and $V=1$ (c) (green). $V=0$ (d) (red) corresponds to a point.}
\label{Triangle_1}
\end{figure}

\begin{figure}
\includegraphics[width=0.45\textwidth]{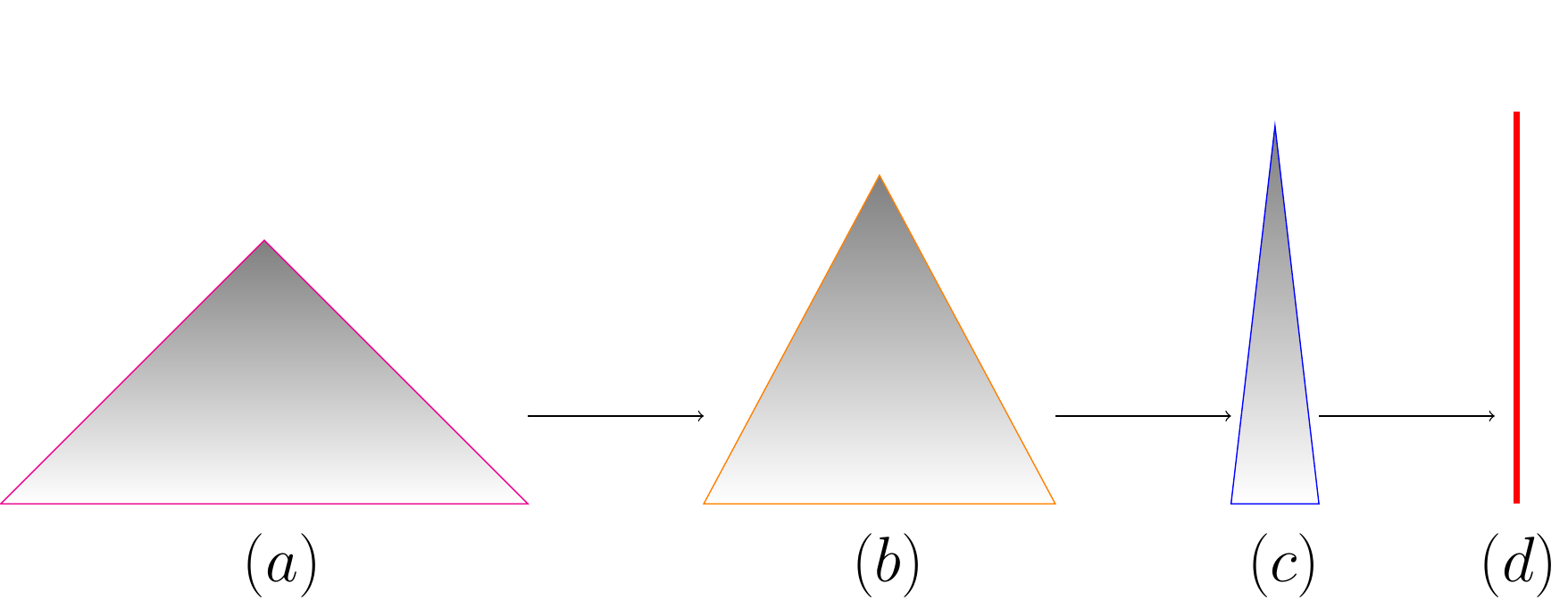}
\caption{(Color online) $(a)-(c)$ Particle-hole triangles for values of interaction strength $V=6$ (a) (purple), 
$V=4$ (b) (orange) and $V=2$ (c) (blue). $V=0$ (d) (red) corresponds to a segment.}
\label{Triangle_2}
\end{figure}

As $V$ decreases from $\infty$, we see three regimes. Up to $V\approx 4$, 
nothing much happens. The particle triangles then start shrinking and shrink
to points at $V=0$. The particle-hole triangles change shape at $V\approx 2$
and become isosceles triangles, they then shrink to segments at $V=0$. This
behaviour is illustrated in Fig.~(\ref{Triangle_1}) and Fig.~(\ref{Triangle_2})
respectively.

\section{Discussion and conclusion}
\label{dc}

To summarise our results, we have given a definition of quantum distances
between pairs of points in the spectral parameter space. We proved that our
definition satisfies the triangle inequalities.  The spectral parameters are
completely general, they could be quasi-momenta, positons labelling Wannier
orbitals, parameters labelling the eigenfunctions of some confining potential
like in a quantum dot or an optical trap.

Our definition of the quantum distances is a purely kinematic one, since it is
in terms of the expectation values of the exchange operators. Thus, if the 
state being considered is the ground state of a system, then the geometry 
defined is manifestly a ground state property. This is in contrast with 
definitions in terms of Green's functions which is a dynamic quantity.

Because of this, our definition can be applied to any state, not necessarily
the ground state. Thus, it could potentially find applications in quantum 
dynamical systems and provide a dynamical geometrical description.

We have applied our formalism to compute and study the distance matrix of the
ground state of the one-dimensional $t-V$ model.  The finite system that we are
studying does not have a phase transition but only a crossover from the
metallic to the insulating regime as $V$ is increased. We observe that the
metallic regime is characterized by a clustering of the distances, either very
small or close to 1. They also show signals of sharp Fermi points. As $V$
increases the distances spread and the Fermi points are washed out.

We have illustrated this behaviour in three ways. 
\begin{itemize}
\item
By examining the 
distances from a fixed point (chosen to be $k=-\pi$) to all the others. This
shows very sharp changes at the Fermi points at low $V$, which smoothen out
at large $V$. 
\item
By examining the nearest neighbour distances and 
constructing a representation of these on a unit circle. 
This representation clearly shows clustering at small $V$ which gets washed 
out at large $V$.
\item
By examining the triangles formed by the distances between three quasi momenta.
The triangles are of two types, both have finite areas in the insulating 
regime which drastically reduce in the metallic regime.
\end{itemize}

In all the three cases discussed above the crossover happens around
$V=2-4$. Since previous studies \cite{Shankar} have established that the
metal-insulator transition occurs at $V=2 $, we conclude that
the ``clustering-declustering" feature that we observe in the distance
matrix is indeed characterizing the metal-insulator crossover.

Our work opens up many directions for further work. One direction is the
following. In this paper we have shown that the distance matrix shows clear
signals of the metal-insulator transition.  There is a large body of
mathematical literature on distance matrices and the geometry of the embedding
space. So the question is, what geometric quantity constructed out of the
distance matrix best describes the metal-insulator transition? We will be
addressing and reporting on this issue in our following paper.  

Another remaining question is the issue of defining geometric phases associated
with loops in the spectral parameter space. Is it possible for general
correlated states? If so, what is the definition?

\section{Acknowledgements}
We are grateful  to  R. Simon, S. Ghosh, R. Anishetty, G. Date, and  
A. Mishra for  useful  discussions.

\appendix

\section{Reduction to the classical Ptolemy problem for $\alpha=2$}
\label{2euc}

In this appendix, we will show that, for $\alpha=2$, the problem of 
proving the triangle inequalities, reduces to the classical Ptolemy
problem in $3$-dimensional Euclidean space.

To state the problem, we are given four normalized vectors in a Hilbert
space, ${\cal H}$, $\vert\chi_\mu\rangle,~\mu=0,\dots,3$, where, with 
reference to the
notation in Section \ref{tineq}, we have defined
$\vert\psi\rangle\equiv\vert\chi_0\rangle$. 
The six distances between these 
four vectors are given by,
\begin{equation}
\label{6dist}
D_{\mu\nu}=\sqrt{1-\vert\langle\chi_\mu\vert\chi_\nu\rangle\vert^2}.
\end{equation}
We will now prove that we can always find $4$ points in a $3$-dimensional
Euclidean space, $\vec x_\mu$, such that,
\begin{equation}
D_{\mu\nu}=\vert\vec x_\mu-\vec x_\nu\vert.
\end{equation}
This reduces the problem to the classical Ptolemy problem.

We can always find a $4$-dimensional subspace of ${\cal H}$ which contains the
four vectors, $\vert\chi_\mu\rangle$. The physical states, forming the manifold
$CP_3$, are in one-to-one correspondence with the pure state density matrices,
\begin{equation}
\label{rhodef}
\rho_\mu\equiv\vert\chi_\mu\rangle\langle\chi_\mu\vert.
\end{equation}
The distances defined in Eq.~\eqref{6dist} can be expressed as,
\begin{equation}
D_{\mu\nu}=\sqrt{1-{\rm tr}\rho_\mu\rho_\nu}.
\end{equation}
Since $\rho_\mu$ are hermitian, they can be 
expressed as a linear combination, with real coefficients, of the 
identity matrix and the $15$ generators of $SU(4)$ in the fundamental 
representation. We denote them by, $T_\alpha,~\alpha=1,\dots,15$. They 
can always be chosen such that,
\begin{equation}
\label{tprops}
{\rm tr}T_\alpha=0,~~~~~~{\rm tr}T_\alpha T_\beta=\delta_{\alpha\beta}.
\end{equation}
Thus we have,
\begin{eqnarray}
\label{rhomuexp}
\rho_\mu&=&a_0I+\sum_{\alpha=1}^{15}a_{\mu}^\alpha T_\alpha\\
a_0&=&\frac{1}{4}{\rm tr}\rho_\mu\\
a_\mu^\alpha&=&{\rm tr}T_\alpha\rho_\mu.
\end{eqnarray}
The fact that, ${\rm tr}\rho_\mu^2={\rm tr}\rho_\mu=1$ implies that,
\begin{equation}
\label{anorm}
a_0=\frac{1}{4},~\vec a_\mu\cdot\vec a_\mu\equiv
\sum_{\alpha=1}^{15}~a_\mu^\alpha a_\mu^\alpha=1-\frac{1}{16}=\frac{15}{16}.
\end{equation}
Note that $\rho_\mu^2=\rho_\mu $ implies other constraints on $\vec a$, 
but these are not relevant for our proof.

Thus, we have shown that each of the physical states, $\rho_\mu$, can be 
represented by a point on a $14$-dimensional sphere of radius 
$\frac{\sqrt{15}}{4}$.

The distance $D_{\mu\nu}$ can be expressed in terms of $\vec a_\mu$,
\begin{eqnarray}
\nonumber
D^2_{\mu\nu}&=&1-{\rm tr}\rho_\mu\rho_\nu\\
\nonumber
&=&\frac{15}{16}-\vec a_\mu\cdot\vec a_\nu\\
\label{dmunuamuanu}
&=&\frac{1}{2}\vert \vec a_\mu-\vec a_\nu\vert^2.
\end{eqnarray}
Thus, if we define $\vec x_\mu\equiv\frac{1}{\sqrt 2}\vec a_\mu$, then we 
have constructed four points, $\vec x_\mu$, in a $15$-dimensional Euclidean
space such that the $6$ distances between them are $D_{\mu\nu}$. Namely,
\begin{equation}
\label{dmunubmubnu}
D_{\mu\nu}^{2}=\vert \vec x_\mu-\vec x_\nu\vert^2.
\end{equation}

We can always find a $3$-dimensional subspace of this $15$-dimensional Euclidean space that contains the four points $\vec x_\mu$.

Hence, we have found $4$ points, $\vec x_\mu$ in a $3$-dimensional
Euclidean vector space such that the $6$ distances between them is $D_{\mu\nu}$.
The problem thus reduces to the classical Ptolemy problem.


%

\end{document}